\documentstyle[12pt]{article}
\begin{document}
\centerline{\bf{Lattice Green Function (at 0) for the 4d }}
\centerline{\bf{Hypercubic Lattice}}
\vskip .2in
\centerline{ M.L. Glasser}
\centerline{Department of Physics}
\centerline{Clarkson University}
\centerline{Potsdam, NY 13699-5820 (USA)}
\vskip .1in
\centerline{and}
\vskip .1in
\centerline{A.J. Guttmann}
\centerline{Department of Mathematics}
\centerline{The University of Melbourne}
\centerline{Parkville, Victoria 3052, Australia}
\vskip .2in

\begin{quotation} The generating function for recurrent Polya walks on the
four dimensional hypercubic lattice is expressed as a Kamp\'{e}-de-F\'{e}riet
function. Various properties of the associated walks are enumerated.
\end{quotation}

Lattice statistics play an important role in many areas of chemistry and
statistical physics. Random lattice walks have been investigated
extensively, both numerically and analytically, and explicit analytic
expressions for recurrent lattice walks (those that end where they began)
are available for several lattices in one, two and three dimensions [1,2].
The interest in this problem follows from the numerous connections between
random walks and other problems in mathematical physics. They form the
basis of certain proofs in the problem of self-avoiding walks [3], they are a
 key
component in the study of lattice dynamics [4], and they have an extensive
following in the mathematical literature [1,3]. Furthermore, d-dimensional
lattice Green functions have recently [5] been shown to underlie the theory
of d-dimensional staircase polygons and to be related to the generating
function for d-dimensional
multinomial coefficients.

The aim of this note is to extend this list to four dimensions by
summarizing a study of the lattice Green function
$$P(z)=P_4(z)=\frac{1}{\pi^4}\int\int\int\int_0^{\pi}
\frac{d^4k}{1-\frac{z}{4}\sum_{j=1}^4\cos(k_j)},\quad\quad\quad |z|\le1
\eqno(1)$$
which generates the statistics for recurrent Polya walks ( uniform step
probabilities) on the 4d hypercubic lattice.

 From (1) it is clear that $P(z)$ is an analytic function of $z^2$ within the
unit circle and has a branch point at $z^2=1$. It is representable as a
single integral in a variety of ways. From the relations in [6], for
example, we have
$$P(z)=\int_0^{\infty}e^{-x}I_0^4(\frac{1}{4}xz)dx \eqno(2)$$
and
$$P(z)=\frac{8}{\pi^3}\int_0^1\frac{{\bf{K}}(k_+){\bf{K}}(k_-)}{
\sqrt{1-x^2}}dx \eqno(3)$$
where
$$k_{\pm}^2=\frac{1}{2}[1\pm x^2z^2\sqrt{1-\frac{1}{4}x^2z^2}-(1-\frac{1}{2}x^2
z^2)\sqrt{1- x^2z^2}] \eqno(4)$$
and {\bf{K}}  denotes the complete elliptic integral of the first kind.

In terms of the variables $u=z^2$ and $p(u)=P(z)$,
$P(z)$ satisfies the Fuchsian differential equation
$$4u^3(u-4)(u-1)p^{(iv)}(u)+8u^2(5u^2-20u+12)p^{\prime\prime\prime}(u)
+u(99u^2-293u+112)p^{\prime\prime}(u)$$
$$+(57u^2-106u+16)p^{\prime}(u)+(3u-2)p(u)=0 \eqno(5)$$
Asymptotic analysis of (2) and  series analysis shows that near $z^2=1$
$$P(z)\cong C_0+C_1(1-z^2)\ln(1-z^2)$$
$$C_0=1.23946712\dots\eqno(6)$$
$$C_1=2/\pi^2$$

Upon expanding (1) in powers of $z^2$, one has
$$P(z)=\sum_{n=0}^{\infty}a_nz^{2n}\eqno(7)$$
where
$$a_n=\frac{1}{\pi^4}\int\int\int\int_0^{\pi}[\frac{1}{4}
\sum_{j=1}^4\cos(k_j)]^{2n}d^4k\eqno(8)$$
On the other hand, by expanding the Bessel function in(2), integrating
term-by-term, and proceeding as in the appendix in [6], we find
$$a_n=[\frac{(1/2)_n}{n!}]^3\;_4F_3[\begin{array}{cccc}
-n/2,&(1-n)/2,&-n,&1/2;\\
1/2-n,&1/2-n,&1;&
\end{array}
1].\eqno(9)$$
Next, by using the identity [6]
$$\;_4F_3(-n,(1-n)/2,-n/2;1/2-n,1/2-n,1;1)=$$
 $$\frac{(1/2)_n}{n!}\;_4F_3(-n,1/2,
1/2,1/2;1/2-n,1/2-n,1/2-n;1)\eqno(10)$$
expressing the latter $\;_4F_3$ in terms of its series, and noting that
$$(-n)_k=(-1)^kn!/(n-k)!,\;\;\quad\quad(1/2-n)_k=(-1)^k(1/2)_n/(1/2)_{n-k},
\eqno(11)$$
we arrive at the double hypergeometric series
$$P(z)=\sum_{m,n=0}^{\infty}\frac{(1/2)_{m+n}(1/2)_n^3(1/2)_m^3}{(1)_{m+n}^3}
\frac{z^{2m}z^{2n}}{n!m!}\eqno(12)$$
which defines a  Kamp\'{e}-de F\'{e}riet function [7]:
$$P(z)=F^{1:3:3}_{3:0:0}[\begin{array}{ccc}
1,1,1;&-;&-;
\end{array} z^2,z^2]\eqno(13)$$
$$=F^{2:1:1}_{0:2:2}[\begin{array}{ccc}
1/2,1;&1/2;&1/2\\
-;&1,1;&1,1;
\end{array} z^2/4,z^2/4].$$

The quantity $a_n$ in (8) is the probability that the random walker ends at
its starting point in 2n steps (not necessarily for the first time). The
number of such walks is therefore $r_n=8^{2n}a_n$, which obeys the recursion
relation
$$n^4r_n-4(20n^4-40n^3+33n^2-13n+2)r_{n-1}$$
$$+256(4n^4-16n^3+23n^2-14n+3)r_{n-2}=0.\eqno(14)$$
The first ten values are given in the table.

\centerline{Number of 2n-step recurrent walks}
\vskip .1in

$$\begin{array}{cc}
n&r_n\\
1&8\\
2&168\\
3&5120\\
4&190120\\
5&7939008\\
6&357713664\\
7&16993726464\\
8&839358285480\\
9&42714450658880\\
10&2225741588095168
\end{array}$$

For large n, for the d-dimensional hypercubic lattice  [8]
$$r_n^{(d)}\sim 2(2d)^{2n}(d/4\pi n)^{d/2}$$
so in the present case $a_n\sim 2/\pi^2n^2$ and $r_n\sim 64^n(2/n^2\pi^2)$.
(For example, this yields $a_{500}\sim 8.10569\times 10^{-7}$, compared to
the exact value $a_{500}=8.0976\times 10^{-7}$, a difference of less than
0.1\%).

To our knowledge, the only previous studies of $P_4(z)$ were by Bender et
al.[8], who examined $P_d(1) $ as a function of $d$ and by a $1/d$ expansion
obtained a value consistent with (6),and by Mayer and Nuttall [9], who derived
a differential equation consistent
with (5). We have not examined $P(z)$ outside the unit circle, although it
is clearly analytic in the $z^2$ plane cut
along the real axis from $z^2=1$ to infinity. While it is doubtful that
$P(z)$ can be expressed in terms of elliptic integrals as for the three
dimensional cubic lattices [2], it is hoped that further analysis of (13)
will
lead to a more transparent representation of its analytic properties.

 \vskip .1in
 \centerline{Acknowlegements}
 MLG expresses his thanks to the University of Melbourne for hospitality
 while this work was
 carried out. AJG acknowleges the support of the Australian Research
 Council. We are grateful to Dr. E.D. Krupnikov for information about the
 Kamp{\'{e}}-de-F{\'{e}}riet functions.

 \vfill\eject
 \centerline{References}

\noindent
[1] M.N. Barber and B. Ninham,{\bf{ Random and Restricted Walks:Theory and
Applications}}.(Gordon and Breach, NY 1977).

\noindent
[2] M.L. Glasser and J.I. Zucker, Proc. Nat. Acad. Sci.{\bf{74}}, 1800
(1977).

\noindent
[3] N. Madras and G Slade,{\bf{ The Self-Avoiding Walk}}.(Birkh\"auser, Boston
 1992)

\noindent
[4] G.S. Joyce, Philos. Trans. R. Soc. London {\bf 273}, 583 (1973)

\noindent
[5] A.J. Guttmann and T. Prellberg, Phys. Rev. {\bf E47}, R2233 (1993)

\noindent
[6] M.L. Glasser and E. Montaldi, Phys. Rev. {\bf{E46}}, R2339 (1993).(The
summation indices on the RHS of (A8) should be $n$ and $k$.)

\noindent
[7] H. Exton,{\bf{ Multiple Hypergeometric Functions and Applications}},

(Horwood, Chichester, 1976).

\noindent
[8] C.M. Bender, S. Boettcher, and L. Mead,. J. Math. Phys. {\bf{35}}, 368
(1994).

\noindent
[9] I.L. Mayer and J. Nuttall, J. Phys. A: Math. Gen.{\bf{20}}, 2357
(1987).

\end{document}